\documentclass[aps,prd,preprint,showpacs,nofootinbib,amsmath,amssymb]{revtex4-1}
\usepackage{epsf}
\usepackage{color}
\usepackage{dcolumn}% Align table columns on decimal point
\usepackage{bm}% bold math
\everymath{\displaystyle}

\begin{document} %%%%%%%%%%%%%%%%%%%%%%%%%%%%%%%%%%%%%%%%%%%%%%%%%

\title{Siberian snake-like behavior for an orbital polarization of a beam of twisted (vortex) electrons}

\author{\firstname{Alexander J.}~\surname{Silenko}}
\email{alsilenko@mail.ru} \affiliation{Bogoliubov Laboratory of Theoretical Physics, Joint Institute for Nuclear Research,
Dubna 141980, Russia,\\Research Institute for
Nuclear Problems, Belarusian State University, Minsk 220030, Belarus}

\author{\firstname{Oleg V.}~\surname{Teryaev}}
\email{teryaev@theor.jinr.ru} \affiliation{Bogoliubov Laboratory
of Theoretical Physics, Joint Institute for Nuclear Research,
Dubna 141980, Russia} %,\\
%National Research Nuclear University ``MEPhI'' (Moscow Engineering
%Physics Institute), Kashirskoe Shosse 31, 115409 Moscow, Russia}

%\date{file ``Siberian snake-like.tex", \today}

\begin {abstract}
The orbital polarization of twisted electrons carrying an intrinsic orbital angular momentum is not influenced by field perturbations in arbitrary magnetic fields. This property means an existence of the Siberian snake-like behavior for an orbital polarization of a beam of twisted electrons in cyclotrons with the main magnetic field and magnetic focusing. As a result, the acceleration of twisted electron beams in cyclotrons necessary for their applications in high-energy-physics experiments considerably simplifies.
\end{abstract}

%\pacs{04.20.Cv; 04.62.+v; 03.65.Sq}
\maketitle

%%%%%%%%%%%%%%%%%%%%%%%%%%%%%%%%%%%%%%%%%%%%%%%%%
%\section{Introduction}
%%%%%%%%%%%%%%%%%%%%%%%%%%%%%%%%%%%%%%%%%%%%%%%%%

Twisted (vortex) electrons which existence has been predicted in Ref. \cite{Bliokh2007} are Dirac particles carrying an intrinsic orbital angular momentum (OAM). They have been discovered in 2010 \cite{UTV}. Since such electrons possess large magnetic moments, this discovery has opened new possibilities in the electron microscopy and investigations of magnetic phenomena
(see Refs. \cite{BliokhSOI,Lloyd,LloydPhysRevLett2012,Rusz,Edstrom,imaging,Observation,OriginDemonstration}
and references therein). Twisted electron beams with large intrinsic OAMs (up to 1000$\hbar$)
have been recently obtained \cite{VGRILLO}. Main properties of twisted electrons have been expounded in the reviews \cite{BliokhSOI,Lloyd}. In Refs. \cite{Manipulating,ResonanceTwistedElectrons}, the general form of relativistic classical and quantum-mechanical equations of motion has been obtained for an intrinsic OAM of twisted electrons in electric and magnetic fields. The orbital polarization (the fixed direction of intrinsic OAMs) of twisted electrons obtained experimentally is parallel or antiparallel to their momenta.

It is known that the acceleration of particles may be accompanied by their passing through spin resonances. These resonances may lead to a depolarization of particle beams. In particular, the beam depolarization is very important for polarized protons. To pass them through spin resonances, one uses Siberian snakes. It has been proposed in Refs.  \cite{Manipulating,ResonanceTwistedElectrons} to accelerate polarized (and unpolarized) beams of twisted electrons for their applications in high-energy-physics experiments. The natural question can be put: are there any resonances depolarizing a beam with an orbital polarization? 

We answer this question in the present work. The system of units $\hbar=1, c=1$ is used.

The equation of motion of the intrinsic OAMs ($\bm L$) of twisted electrons or other twisted particles in electric and magnetic fields has the form \cite{Manipulating,ResonanceTwistedElectrons}
\begin{equation}\begin{array}{c}
\frac{d\bm L}{dt}=\bm\Omega\times\bm L,\qquad \bm\Omega=-\frac{e}{2mc\gamma}\left(\bm B-\bm\beta\times\bm E\right),
\end{array} \label{reLarpr} \end{equation} where $\gamma$ is the Lorentz factor and $\bm\beta=\bm V/c$.
The motion of the intrinsic OAMs is the Larmor precession.
Otherwise, the angular velocity of their cyclotron motion is defined by
\begin{equation}\begin{array}{c}
\frac{d\bm N}{dt}=\bm\omega\times\bm N,\qquad \bm\omega=-\frac{e}{mc\gamma}\left(\bm B-\frac{\bm\beta\times\bm E}{\beta}\right),
\end{array} \label{recycpr} \end{equation} where $\bm N=\bm p/p$ is the unit vector along the momentum direction. One usually applies cyclotrons with the main vertical magnetic field and magnetic focusing. Evidently, the accelerating longitudinal electric field does not influence the quantities $\bm\omega$ and $\bm\Omega$. In this case, $\bm\beta\times\bm E$. As a result, \begin{equation}\bm\omega=2\bm\Omega\label{maineqn} \end{equation} for any beam energy. 

Equation (\ref{maineqn}) means that any perturbation of the intrinsic-OAM motion (originated from a horizontal magnetic field at any path section) twice affects the electron during the period of the intrinsic-OAM rotation. It is important that the phases of the intrinsic-OAM oscillation at the moments of the twofold action of the perturbation are opposite and the deflections of the intrinsic OAM caused by this perturbation are also opposite. Thus, the acceleration of twisted electrons in usual cyclotrons with the main vertical magnetic field and magnetic focusing does not lead to the resonant orbital depolarization.  

The result presented means an existence of the Siberian snake-like behavior for an orbital polarization of a beam of twisted electrons. This is the wonderful property of orbitally polarized twisted electron beams significantly simplifying their acceleration in cyclotrons.

The work
was supported in part by the Belarusian Republican
Foundation for Fundamental Research (Grant
No. $\Phi$18D-002), by the Heisenberg-Landau
program of the Federal Ministry of Education and
Research of Germany
(Bundesministerium f\"{u}r Bildung und
Forschung), 
and by the Russian Foundation for Basic Research
(Grant No. 16-02-00844-A).

%\newpage
%\begin{footnotesize}

\end{document}